\documentstyle[aps,preprint,prl]{revtex}

\begin{document}
\bibliographystyle{plain}
\title{Negative modes  and decay-rate  transition}
\author{Soo-Young Lee$^a$, Hungsoo Kim$^b$, D.K.Park$^a$,
Jae Kwan Kim$^b$}
\address{$^a$ Department of Physics, Kyungnam University,
Masan, 631-701, Korea.\\
$^b$ Department of Physics, Korea Advanced Institute of Science and
Technology, \\Taejon, 305-701, Korea.}
\date{\today}
 \maketitle

 \begin{abstract}
We investigate a relationship between the number of the negative modes 
around periodic instanton solution and the type of the decay-rate transition.
It is shown that for the case of first-order decay-rate transition
 the lowest positive mode at low
energy periodic instanton becomes additional negative mode at high energy
regime,
 while in the second-order case there is only one negative mode in the full
range of energy.
  This kind of analysis on the negative modes makes it possible to derive
 the criterion for the first-order transition.
 \end{abstract}
\newpage
\section{Introduction}
Recently, the decay-rate transition have attracted much attention in various
fields from condensed matter physics\cite{bl94} to particle
 physics\cite{ha98,ku97} and 
 cosmology\cite{li83}. This implies that
it is a general phenomenon which appears frequently in nature.

Since the decay rate of metastable state is expressed 
as
\begin{equation}
\Gamma \propto e^{-S[\phi]},
\end{equation}
where $S[\phi]$ is the Euclidean action,
the classical solution in Euclidean space makes a dominant contribution
to the decay rate. At zero temperature the classical localized solution is a 
bounce which has infinite Euclidean period\cite{co77}.
 As temperature increases,
due to thermal assistance a periodic solution called periodic 
instanton\cite{kh91} plays
an important role in the decay rate, and then at higher temperature 
thermal activation becomes dominated, which is described by static 
sphaleron\cite{ma83}. The decay-rate transition means the transition from
the periodic instanton-dominated to the sphaleron-dominated regimes.

The decay-rate transition was first argued to be smooth in quantum 
mechanics by Affleck\cite{af81}. After then, Chudnovsky\cite{ch92}
 has shown that it is
possible for the transition to be sharp first-order as well as smooth 
second-order, and the type of the transition depends on the shape of the
potential barrier. He has also shown that the order of the transition is 
easily conjectured by $\beta$-vs-$E$ plot, where $\beta$ and $E$ are
Euclidean period and energy, respectively. The sharp first-order transition
takes place when $\beta(E)$ curve possesses a minimum at $E=E_c$ lower than
the energy of sphaleron $E_s$. Based on Chudnovsky's observation the sharp 
first-order transitions are found at spin tunneling systems with\cite{le98}
 and without\cite{li98} external magnetic field.

Using Chudnovsky's idea a sufficient criterion for the first-order
transition is obtained by carrying out the nonlinear perturbation near
the sphaleron in two-dimensional string model\cite{go97}. Inspired by 
spin-tunneling problem, this criterion is subsquently extended to the 
quantum mechanical model when mass is position-dependent\cite{mu99}.

Recently, it is conjectured that the number of the negative modes of the
fluctuation operator may change at the bifurcation point\cite{ku97}.
This means it is possible to determine the type of the transition by
counting the number of the negative mode near sphaleron.  
The purpose of the present paper is to address this issue by analyzing
the properties of the negative
mode, especially near bifurcation point, explicitly with a concrete quantum
mechanical example and to derive the criterion for the sharp first-order
or smooth second-order transition in this context.
 As will be shown, the criterion derived from this viewpoint exactly
concides with that of Ref.\cite{go97}, in which the criterion is derived
by computing the variation of the period near sphaleron.
In Sec.II we examine the conjecture of Ref.\cite{ku97} by making use
of the simple quantum mechanical model. It is shown that the change
of the number of negative modes at bifurcation point is really
realized in this model. Getting some physical insight from this simple
model, we explore the mechanism of this phenomena at Sec.III. We will
show that the number of negative modes are conjectured by considering
some relations between the classical solutions which have same period
and different action values.
 Based on the anlysis of Secs.II and III, we derive
the criterion for the first-order transition in Sec.IV, and 
in the final section a brief conclusion will be given.

\section{Number of negative modes and decay-rate transition: 
quantum mechanical example}

In this section we calculate numerically  periodic instantons and eigenvalues
of their fluctuation operators for a quantum mechanical potential
\begin{equation}
V(\phi ) = \frac{4+\alpha}{12} -\frac{1}{2}\phi^2 - \frac{\alpha}{4}\phi^4
+\frac{\alpha+1}{6}\phi^6 -\frac{\gamma}{3} \phi^3,
\label{potential}
\end{equation}
where $\alpha$ and $\gamma$ are some real parameters.
Using the criterion of Ref.\cite{mu99}, it is easily shown that 
 when $\gamma=0$ the transition from classical to quantum tunneling 
is first order
for positive $ \alpha $, and in the range of $-1 < \alpha < 0$
 the second-order transition takes place. 
When $\gamma \neq 0$, in the range of $\gamma^2 < \frac{9}{10}\alpha$ 
the first-order transition occurs, and we find that in the range of
 $\gamma^2 > \frac{9}{10}\alpha$
 there are two types of
second-order transition;
 one is usual second-order and the other is  unusual second-order
transition accompanied by a first-order transition within quantum tunneling
 region.

Since the Euclidean action is
\begin{equation}
S[\phi] = \int_{0}^{\beta} d\tau \,\,\left[ \frac{1}{2} \dot{\phi}^2 + V(\phi)
\right],
\label{action}
\end{equation}
the equation of motion reads
\begin{equation}
\ddot{\phi} = V'(\phi),
\end{equation}
where the prime denotes the derivative with respect to $\phi$,
which implies  energy conservation in Euclidean space
\begin{equation}
\frac{1}{2} \dot{\phi}^2 = V(\phi) - E.
\end{equation}
 From this equation the period is expressed as
\begin{equation}
\beta = 2 \int_{\phi_1(E)}^{\phi_2(E)} \frac{d\phi}{\sqrt{2(V(\phi)-E)}},
\end{equation}
where $\phi_1 (E)$ and $\phi_2 (E)$ are turning points.

Fig. 1 shows  plots of potential $V(\phi)$, period $\beta(E)$, and
 action $S(\beta)$
for the three types of decay-rate transition. One can see that the bifurcation
 points of
periodic instantions correspond to the points of $\frac{d\beta}{dE} =0$. 
The periodic instantons in the range of $\frac{d^2 S}{d \beta^2} > 0$ or ,
 equivalently,
$\frac{d E}{d \beta} > 0$, does not contribute to the decay rate since 
there are other classical solutions with the same period and lower action.

Now consider a small fluctuation from the periodic instanton:
\begin{equation}
\phi(\tau) = \phi_p (\tau)+\eta(\tau).
\end{equation}
Then, the action of Eq.(\ref{action}) becomes upto the second order 
of $\eta(\tau)$
\begin{equation}
S[\phi] = S[\phi_p] + \frac{1}{2} \int_0^{\beta} d\tau \,\, \eta(\tau) \hat{M} \eta(\tau),
\end{equation}
where the fluctuation operator is
\begin{equation}
\hat{M} = -\frac{\partial^2}{\partial \tau^2} + V''(\phi_p).
\end{equation}
Let us consider the eigenvalue equation of $\hat{M}$
\begin{equation}
\hat{M} x_n= \epsilon_n x_n. \,\,\,\,(n=-1,0,1,2,3 \cdots).
\end{equation}
For convenience we fix the zero mode as $x_0$, so that $\epsilon_0 =0$.

The eigenvalues and eigenfunctions of the fluctuation operator $\hat{M}$ 
can be numerically
calculated by the diagonalization method of the matrix mainly 
composed of Fourier coefficients
of $V''(\phi_p)$. Several eigenvalues from the lowest one are plotted 
in Fig. 2 for the three
different types of decay-rate transition.
 It is clear that the zero mode($\dot{\phi_p}(\tau)$)
originated from the time translational
symmetry always exists and at the bifurcation points another zero mode occurs,
 which
is relevant not to the symmetry of system, but to the symmetry in the vicinity of the periodic
instanton at the bifurcation point.
The periodic instantons in the range of $\frac{d^2 S}{d \beta^2} > 0$ have 
two negative
modes, which means that along the directions of the negative modes 
in function space
 the periodic instanton gives maximun action value. 
In the next section we will explore the mechanism on the change of the
negative modes by considering the classical solutions in function space.

\section{ Analysis of negative modes in function space}

In the double well potential like Eq.(\ref{potential}) the action has a 
lower bound in the
function space with  a fixed period $\beta$. Since the periodic instanton has
 one or two
negative modes, it is not a minimum in the function space but a saddle point 
having
one or two negative mode directions along which the action value decreases 
approaching
other saddle point or some minimum. Fortunately in this quantum mechanical
 model one can
find all possible saddle points and minima at a given period $\beta$ 
all of which are
solutions of Euler-Lagrange equation.
Therefore, it is possible to expect what solution one can meet ultimately 
if one follows a particular negative mode direction.

The number of negative modes around spaleron depends on the period $\beta$.
This fact is easily shown as follows. The curvature of barrier top has a relation
\begin{equation}
V''(\phi_s) = 2\pi/\beta_s,
\end{equation}
where $\beta_s$ is the period with which the periodic instanton meets
 the sphaleron.
Then the eigenvalue equation for the fluctuation is
\begin{equation}
(-\frac{d^2}{d\tau^2} - (\frac{2\pi}{\beta_s})^2 ) \chi_n(\tau) = \lambda_n \chi_n (\tau)
\end{equation}
where
\begin{eqnarray}
\chi_n (\tau) &=& \cos (\frac{2\pi n}{\beta} \tau), \,\,\sin (\frac{2\pi n}{\beta} \tau)  \\
\lambda_n &=& (\frac{2\pi n}{\beta})^2 - (\frac{2\pi}{\beta_s})^2, \,\,\, (n= 0, 1,2, \cdots ).
\end{eqnarray}
Except for $n=0$ case, the states are  two-fold degenerated.
The number of negative modes increases with $\beta$; for example,
 for $0< \beta < \beta_s$,
there is only one negative mode, and for $\beta_s < \beta < 2\beta_s$
 three negative modes and so on.
The increase of  negative modes will be shown to be natural in the followings.

Let us consider the sphaleron solution $\phi_s$ with a period 
$\beta_1 < \beta_s$  
as shown in Fig.1 (c). Note that at this period there is no any other 
non-trivial solution. 
Two trivial stable solutions are $\phi_+$ and $\phi_-$ staying at well minima,
 respectively(see Fig.1 (a)).  
The sphaleron has one negative mode
 and infinite number of positive modes which are two-fold degenerated.
In Fig.3 all solutions with period $\beta_1$ and several
lowest eigenfunctions are described.
It is clear that the sphaleron solution can go to $\phi_+$ or $\phi_-$ 
by adding
or subtracting the negative eigenfunction $x_{-1}$. So, this situation can be
 described
graphically in function space as Fig.4. It is certain that one classical
 solution can flow along  
the negative mode directions into the other solutions with lower action or to
the vicinity of those.

Now, consider the periodic instanton $\phi_p$ with a period 
$\beta_2 > \beta_s$ in Fig.1(c).
Fig.5 shows all possible solutions with period $\beta_2$ and several eigenfuctions
of the fluctuation operator at the periodic instantion. The periodic instanton has
one negative mode($x_{-1}$), one zero mode($x_0$), and infinite number of
 positive modes($x_n$, $n=1,2,3,\cdots$).
Since the sphaleron solution $\phi_s$ flows along the negative mode 
direction($x_{-1}$) to
$\phi_+$ or $\phi_-$,  this periodic instanton can not reach  exactly 
$\phi_+$ or $\phi_-$
along the negative mode direction, but  go through the vicinity of those stable trivial
solutions. By the zero mode the periodic instanton is shifted without any variation of
action value, which  is the time translational symmetry. The lowest positive mode $x_1$
connects $\phi_p$ with the sphaleron solution and the $\beta_2/2-$shifted periodic instanton
 $\phi^*_p$.
Combining all these facts one can summarize the situation as Fig.6.
From this figure it is clear that the sphaleron with period $\beta_2$ has three negative modes;
one is  $x_{-1}$ which connects $\phi_s$ with $\phi_+$ and $\phi_-$ like Fig.4 (a) and 
the others are $x_0$ and $x_1$ which connect with the $\beta_2/4-$shifted periodic instanton
$\phi^{**}_p$ and $\phi_p$, respectively.
The sphaleron with the period of $2\beta_s < \beta < 3\beta_s$ has five negative modes.
This is explained by considering the solution composed of two  periodic instanton with period
$\beta/2$ as shown in Fig.7. Thus, the additional two negative mode $x_2$ and $x_3$ are the
direction to the periodic instanton with half period, $\phi_{p,\beta/2}$ and
the $\beta/4-$shifted half periodic instanton $\phi^{**}_{p,\beta/2}$
as shown Fig.7.

In the case of first-order transition the periodic instanton can have two negative modes.
Let us consider the periodic instanton $\phi_{p,u}$ with period $\beta_3$ (see Fig.1 (f)) 
which has two negative modes
$x_{-1}$ and $x_1$,  one zero mode $x_0$, and infinite number of positive modes.
In Fig.8 (a) all possible solutions with $\beta=\beta_s$ are drawn. 
All non-trivial solutions can be connected along
$x_1$ direction, which is shown in Fig.8 (b). This figure shows that $\phi_{p,u}$ has negative
eigenvalue  and $\phi_s$ and $\phi_{p,d}$ have positive eigenvalue in $x_1$  direction.
As $\beta_3$ decreases, the $\phi_{p,u}$ and $\phi_{p,d}$ approach each other, and
 at the bifurcation point they have the same action value and then disppear.
 When $\beta_3$ increases through $\beta_s$, $\phi_s$ and $\phi_{p,u}$ are merging each other
and then creating a maximum $\phi_s$.

Finally, consider the case of unusual second-order transition.
All possible solutions with period $\beta_4$(see Fig.1 (i))  and their positions 
along $x_1$ direction 
in function space are described in Fig.9. 
In this case the $\phi_{p,u}$ and $\phi_{p,d}$ merge each other at bifurcation point $\beta_{c1}$
and $\phi_{p,m}$ and $\phi_{p,u}$ do at bifurcation point $\beta_{c2}$, 
and  $\phi_s$ has negative eigenvalues in  $x_0$ and $x_1$ 
directions as well as in $x_{-1}$ direction.

From this analysis we can conclude that additional negative modes require
existence of other solutions with lower action value, which is equivalent
to the situation that occurrence of new maximum always accompanies new minima.

\section{Criterion of first-order transition}
In this section we derive the criterion for the first-order transition
from the fact that in the first-order transition the periodic instantion near sphaleron has
two negative modes while in the second-order case it can have only one negative mode.
This fact can be applied to the field theoretical as well as quantum mechanical models.
In this paper we will concentrate on scalar field theories. However, the extensions to 
the other kinds of field theories such as non-linear $O(3)$ model with\cite{pi94}
and without\cite{mo89} Skyrme term are straightforward, which will be addressed in the 
separate publication.

We start with Euclidean action for a scalar field $\phi$
\begin{equation}
S[\phi] = \int_0^{\beta} d\tau \,\int d\vec{x}\,\,[ \frac{1}{2} \dot{\phi}^2 +\frac{1}{2} ( \nabla \phi)^2
+ V(\phi) ].
\end{equation}
Consider a small fluctuation $\eta (\vec{x},\tau )$ from a periodic instantion
$\phi_p (\vec{x},\tau)$, then the action is
\begin{equation}
S[\phi ] = S[\phi_p] + \frac{1}{2} \int_0^{\beta} \int d\vec{x} \,\,
\eta (\vec{x},\tau ) \hat{M} \eta (\vec{x},\tau ),
\end{equation}
where
\begin{equation}
\hat{M} = -\frac{\partial^2}{\partial \tau^2} - \nabla^2 + V''(\phi_p).
\end{equation}
Since the periodic instanton near the sphaleron can be expressed as\cite{go97}
\begin{equation}
\phi_p (\vec{x},\tau ) = \phi_s (\vec{x}) + a g_0(\vec{x})\cos \omega \tau
+ a^2 ( g_1 (\vec{x}) + g_2 (\vec{x}) \cos 2\omega \tau ),
\end{equation}
where $a$ is a small amplitude and $g_0(\vec{x})$,$g_1(\vec{x})$, and 
$g_2(\vec{x})$ will be determined shortly,
the eigenvalue equation becomes upto the order of $a^2$
\begin{equation}
(-\hat{l} + \hat{h} + \hat{H}' ) \eta (\vec{x},\tau) = \epsilon 
\eta (\vec{x},\tau),
\end{equation}
where
\begin{eqnarray}
\hat{l} &=& \frac{\partial^2}{\partial \tau^2},\nonumber \\
\hat{h} &=& - \nabla^2 + V''(\phi_s),     \\
\hat{H}'&=& aV'''(\phi_s) g_0(\vec{x}) \cos \omega \tau + a^2 [ V'''(\phi_s)
(g_1(\vec{x}) + g_2(\vec{x}) \cos 2 \omega \tau ) \nonumber \\ 
& &+ \frac{1}{2}V''''(\phi_s)
g_0^2(\vec{x}) \cos^2 \omega \tau ]. \nonumber
\end{eqnarray}
We treat the operator $(-\hat{l}+\hat{h})$ as an unperturbed part.
The eigenspectrum of $\hat{l}$ is
\begin{equation}
\hat{l} (|\sin n \omega \tau >, |\cos n \omega \tau > ) =
-(n \omega)^2 (|\sin n\omega \tau >, |\cos n \omega \tau > ).
\end{equation}
Let us assume the eigenspectrum of $\hat{h}$ as
\begin{equation}
\hat{h} | u_n (\vec{x}) > = \alpha_n |u_n (\vec{x}) >, \,\,\, (n = 0,1,2,\cdots ).
\end{equation}
The $\hat{h}$ operator has only one negative mode $|u_0(\vec{x})>$. From the
fact that $\dot{\phi_p}$ and $\partial \phi_p/\partial x_i$ are zero modes
due to the time and space translational symmetry, one can determine
$\omega^2 = - \alpha_0$ and
\begin{eqnarray}
g_0(\vec{x}) &=& u_0 (\vec{x}), \nonumber \\
g_1(\vec{x}) &=& -\frac{1}{4} \hat{h}^{-1} u_0^2(\vec{x}) V''' (\phi_s), \\
g_2(\vec{x}) &=& -\frac{1}{4} (\hat{h} + 4 \omega^2)^{-1} u_0^2 (\vec{x})
V'''(\phi_s). \nonumber
\end{eqnarray}
The zero modes of the unperturbed part or the fluctuation operator at
sphaleron are $|u_0 \sin \omega \tau >$, $|u_0 \cos \omega \tau >$, and
$|u_m >, \,\, (m = 1,2,\cdots D)$. When the perturbation $\hat{H}'$ turns on,
only the second zero mode $| u_0 \cos \omega \tau >$ can be shifted from zero 
 while other zero modes
does not affected by the perturbation because these zero modes are relavant
to system's symmetries, i.e., the time and space translational symmetries.

Therefore, using the standard perturbation theory, we can determine the 
number of negative modes in the vicinity of sphaleron. Let $\epsilon_0$ and
$\epsilon_1$ be the perturbed eigenvalues of $|u_0\sin \omega \tau >$ and
$|u_0 \cos \omega \tau >$, respectively. Then, when 
$\epsilon_1 - \epsilon_0 < 0$ the number of negative mode becomes two, so that
the transition is first order. The criterion for the first-order transition
is, then, straightforwardly derived as
\begin{equation}
<u_0 | g_2 V''' +\frac{1}{4} V'''' g_0^2 |u_0 > 
- \frac{1}{2} < u_0 |g_0 V''' \hat{h}^{-1}g_0 V''' | u_0 > \,\, < \,0.
\label{cri}
\end{equation}
One can easily show that Eq.(\ref{cri}) coincides with the criterion for 
the first-order transition in Ref.\cite{go97} which is obtained from
a different point of view.

\section{Conclusion}
We investigate a relationship between the number of the negative modes
around periodic instanton solution and the type of the decay-rate transition.
It is shown explicitly by calculating the several eigenvalues numerically
in quantum mechanical model that the number of the negative modes 
is changed  at the bifurcation points. 
The mechanism of this phenomenon in the various types of phase diagram
is discussed at  Sec.III by considering the projection of the compact
action manifold in function space.
This kind of analysis makes it possible to derive the criterion for the
sharp first-order or smooth second-order transition from classical to 
quantum tunneling regimes. Although our derivation of criterion is 
exactly same with that of Ref.\cite{go97}, in which the criterion 
is derived from completely different point of view, the viewpoint used
in this paper seems to provide a more profound understanding on the tunneling
phenomena. We hope the relation of negative modes of hessian and the type
of the transition explored in this paper furnishes some physical insight
to understand more complicated tunneling phenomena in electroweak theory
and cosmology.

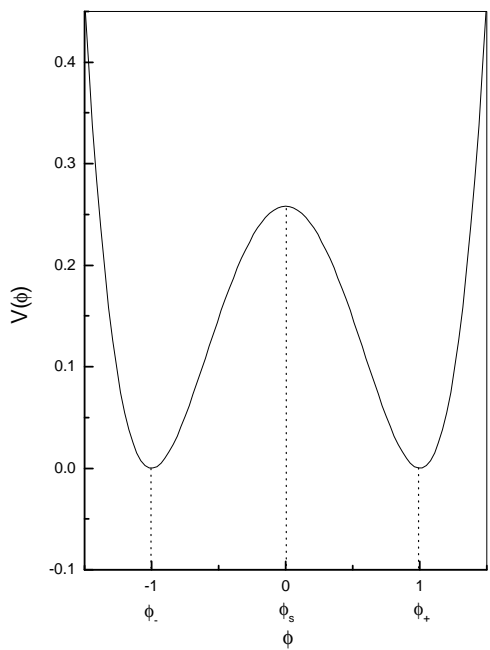
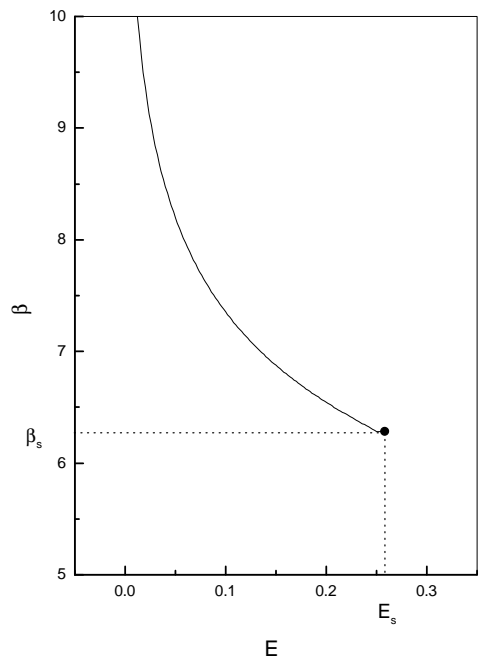
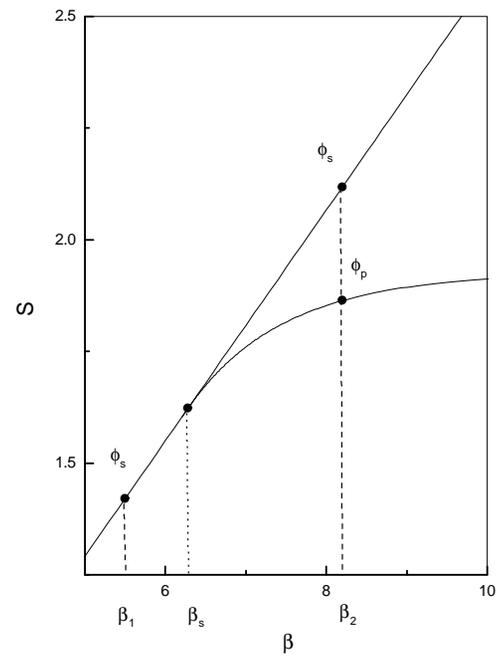
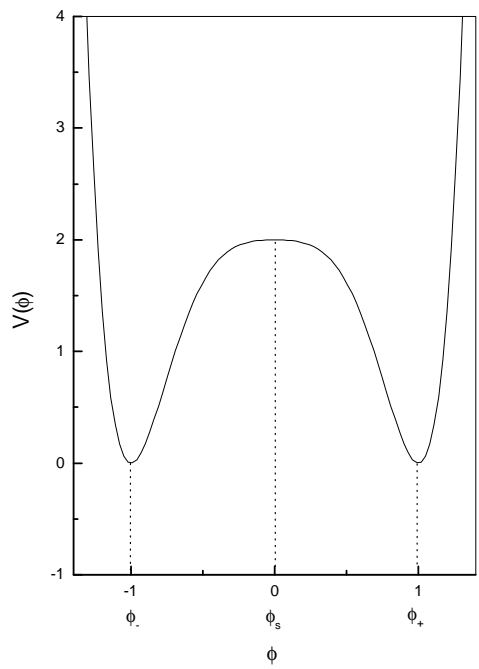
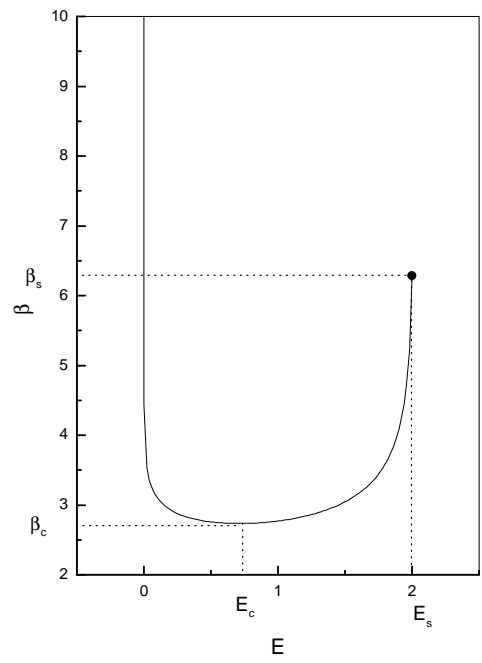
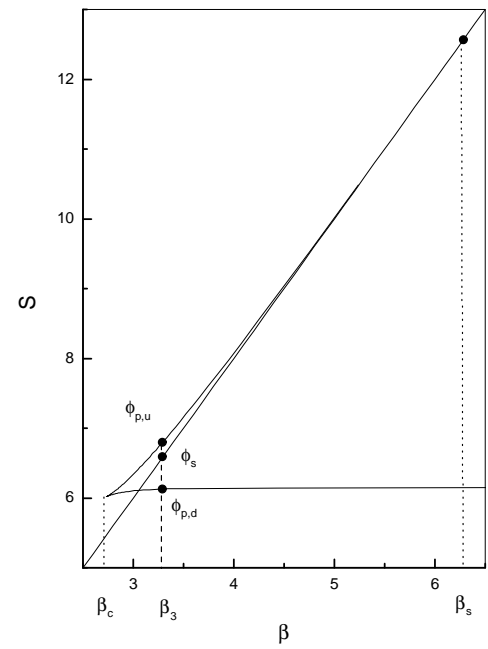
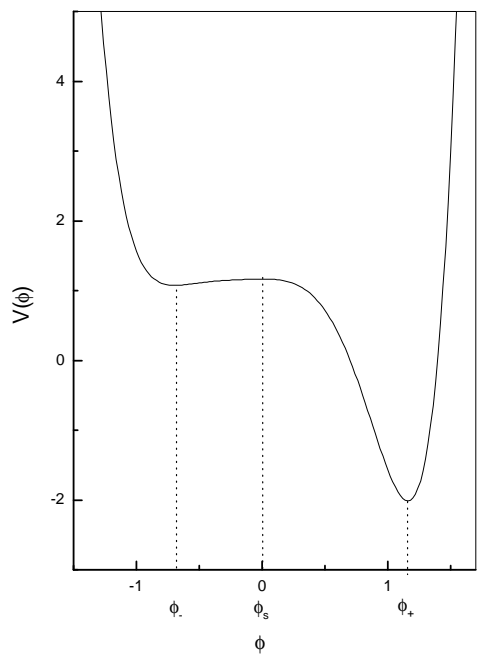
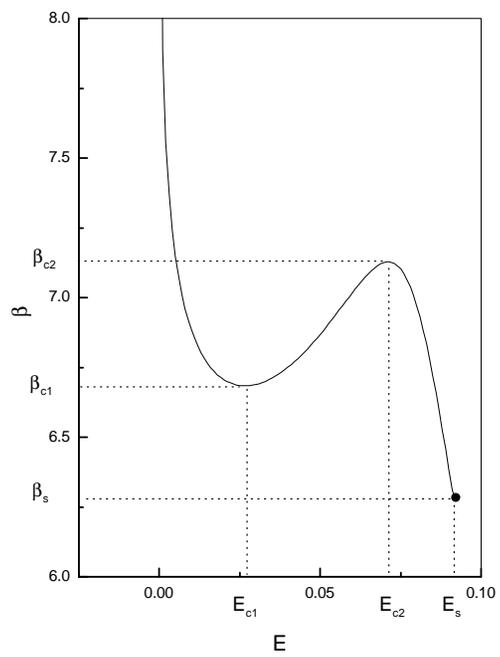
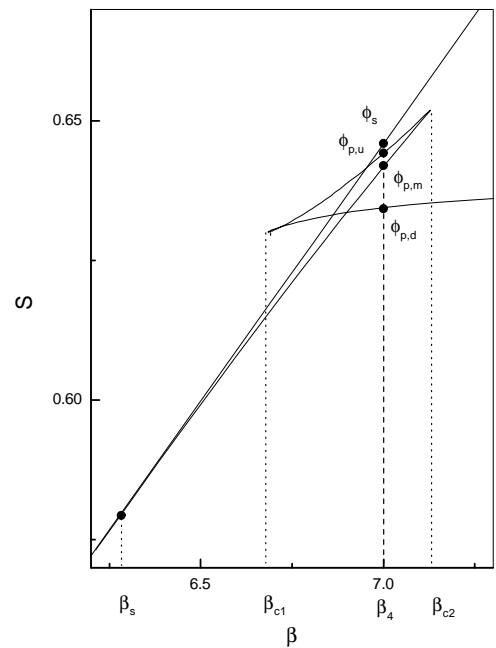
\begin{figure}
\caption{Plots of $V(\phi)$, $\beta(E)$, and $S(\beta)$ for various
types of transition. (a,b,c) usual second order transition at $\alpha =-0.9$ and $\gamma =0$. (d,e,f) first-order transition at $\alpha=20$ and $\gamma=0$.
(g,h,i) unusual second-order transition at $\alpha=10$ and $\gamma=4.7$. In 
this case $\beta(E)$ and $S(\beta)$ are calculated using a potential 
$V(\phi)-V(\phi_-)$.}
\end{figure}
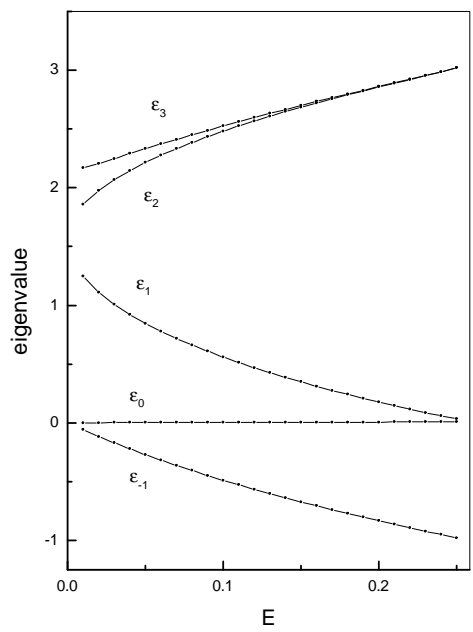
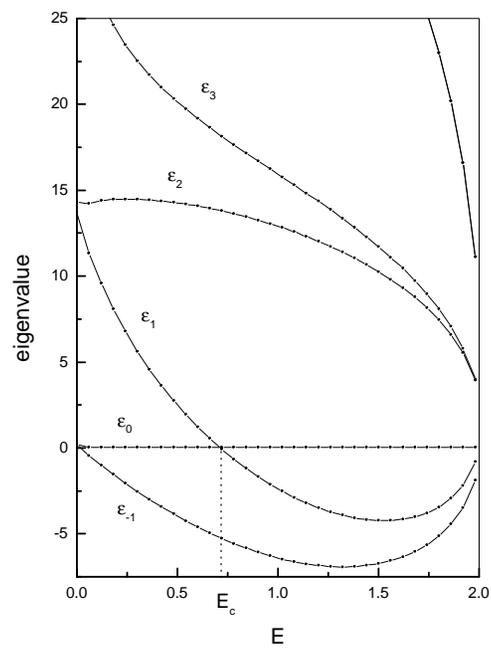
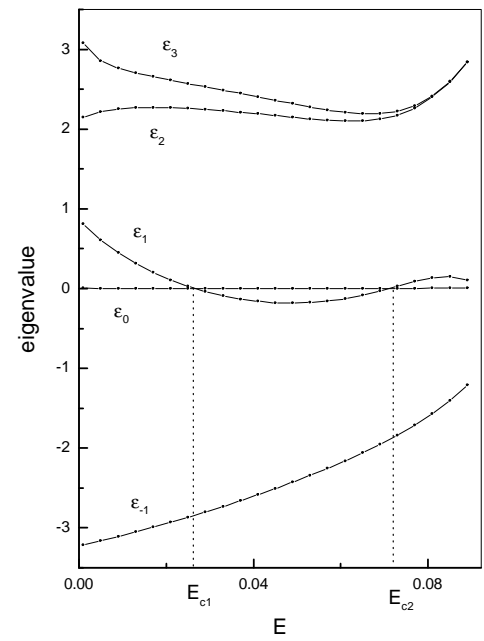
\begin{figure}
\caption{Plots of several lowest eigenvalues of $\hat{M}$. (a) Usual second-order transition
case. There is only one negative mode in the full range of energy. 
(b) First-order transition case. On can see that $\epsilon_1$ becomes additional negative mode when $E$ is larger than $E_c$ which corresponds to the bifurcation point. 
(c) Unusual second-order transition case. One can see that $\epsilon_1$ becomes
additional negative mode at $E_{c1} < E < E_{c2}$. $E_{c1}$ and $E_{c2}$
correspond to two different bifurcation points.  }
\end{figure}
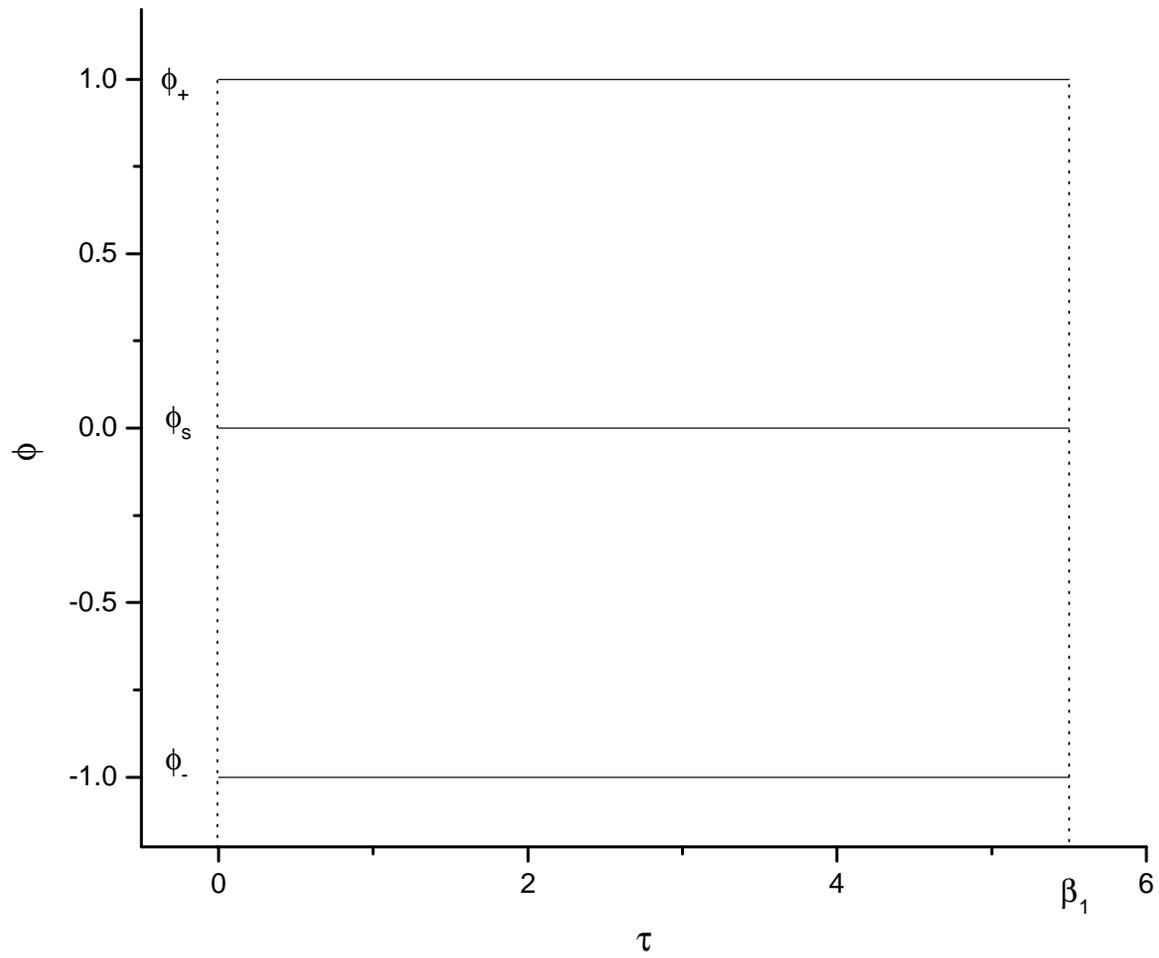
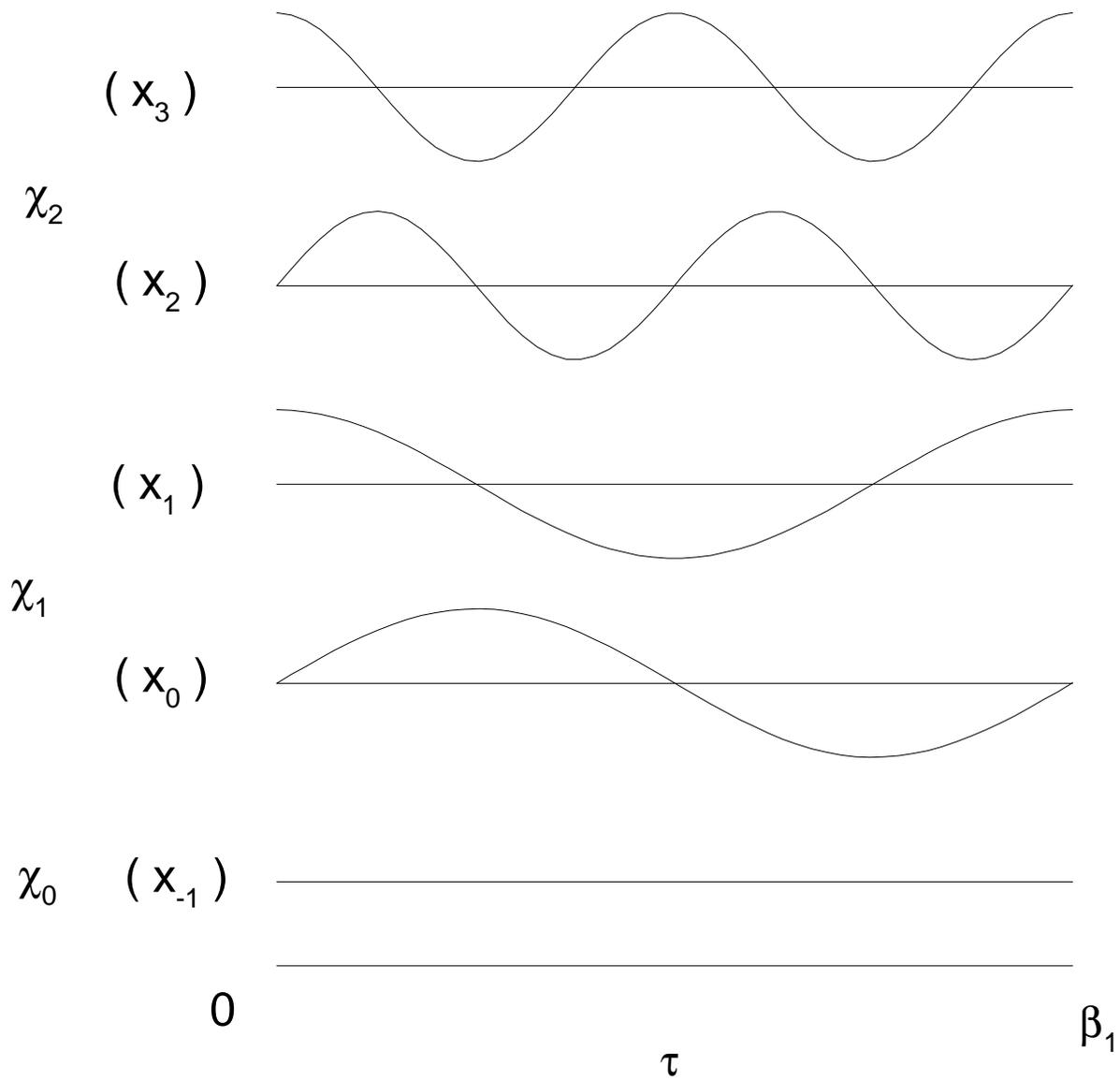
\begin{figure}
\caption{(a) Plots of three trivial solutions sitting at barrier top($\phi_s$)
and bottoms($\phi_+$, $\phi_-$).
(b) Plots of eigenfunctions $x_n\, (n=-1,0,1,2,3)$ around sphaleron solution
when $\beta=\beta_1$ which is smaller than $\beta_s$.} 
\end{figure}
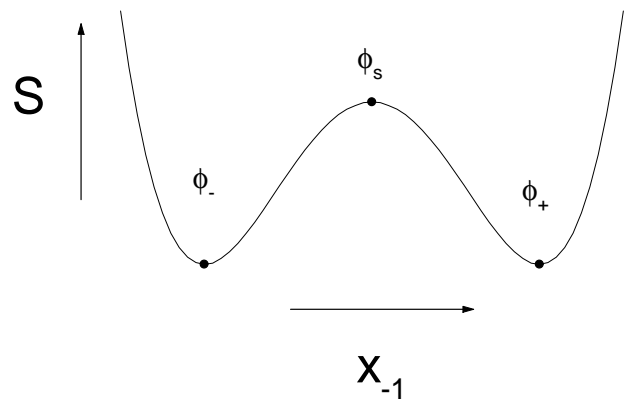
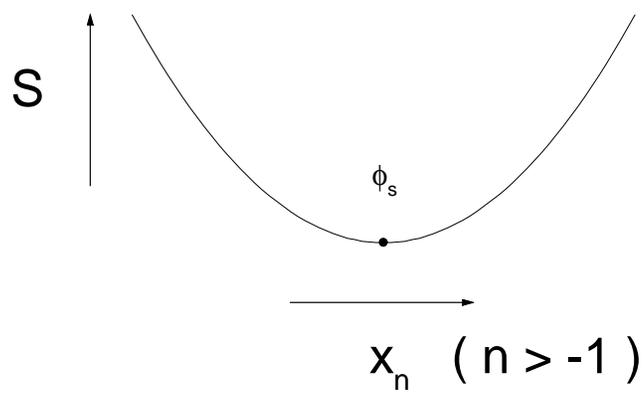
\begin{figure}
\caption{Schematic description of the action value along the (a) negative
and (b) positive mode directions in function space with $\beta=\beta_1$.}
\end{figure}
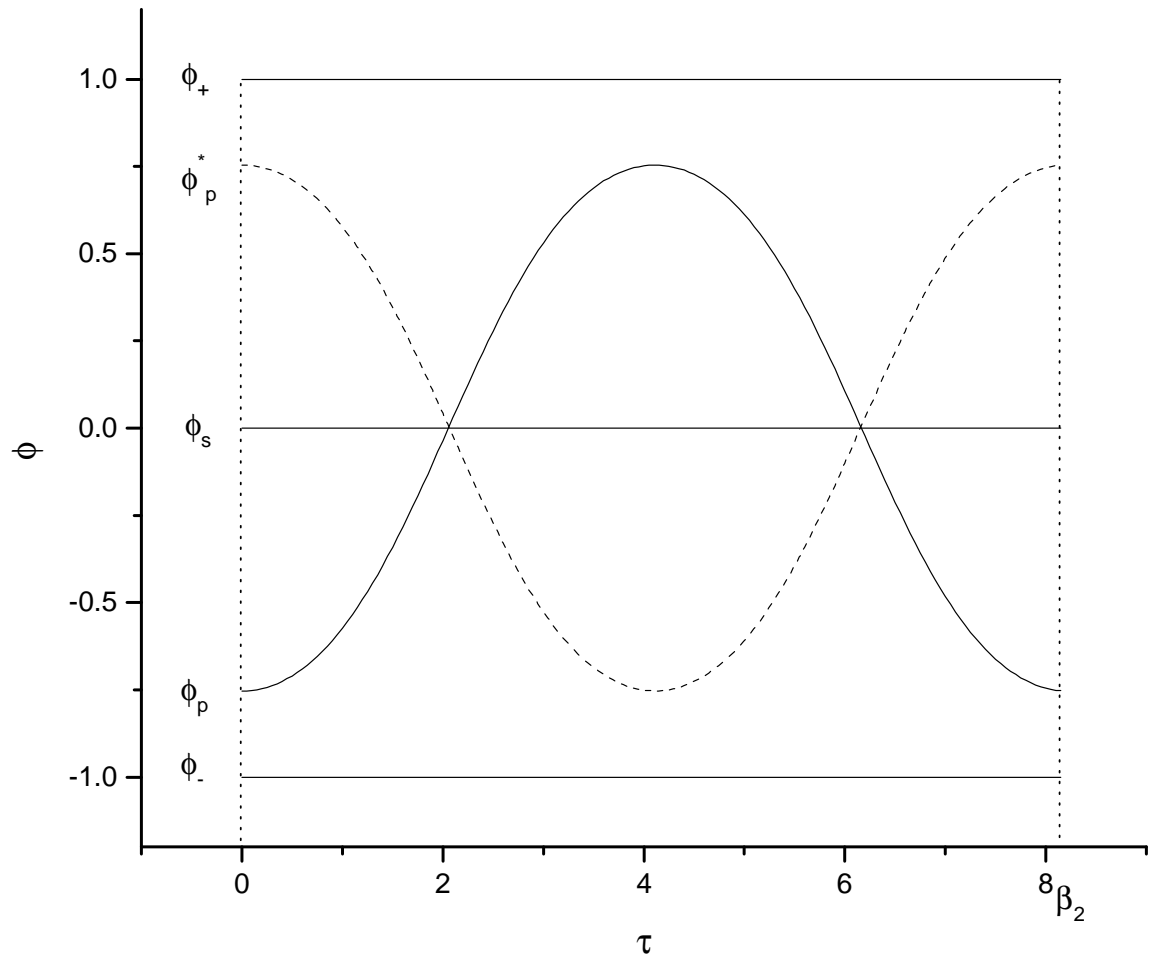
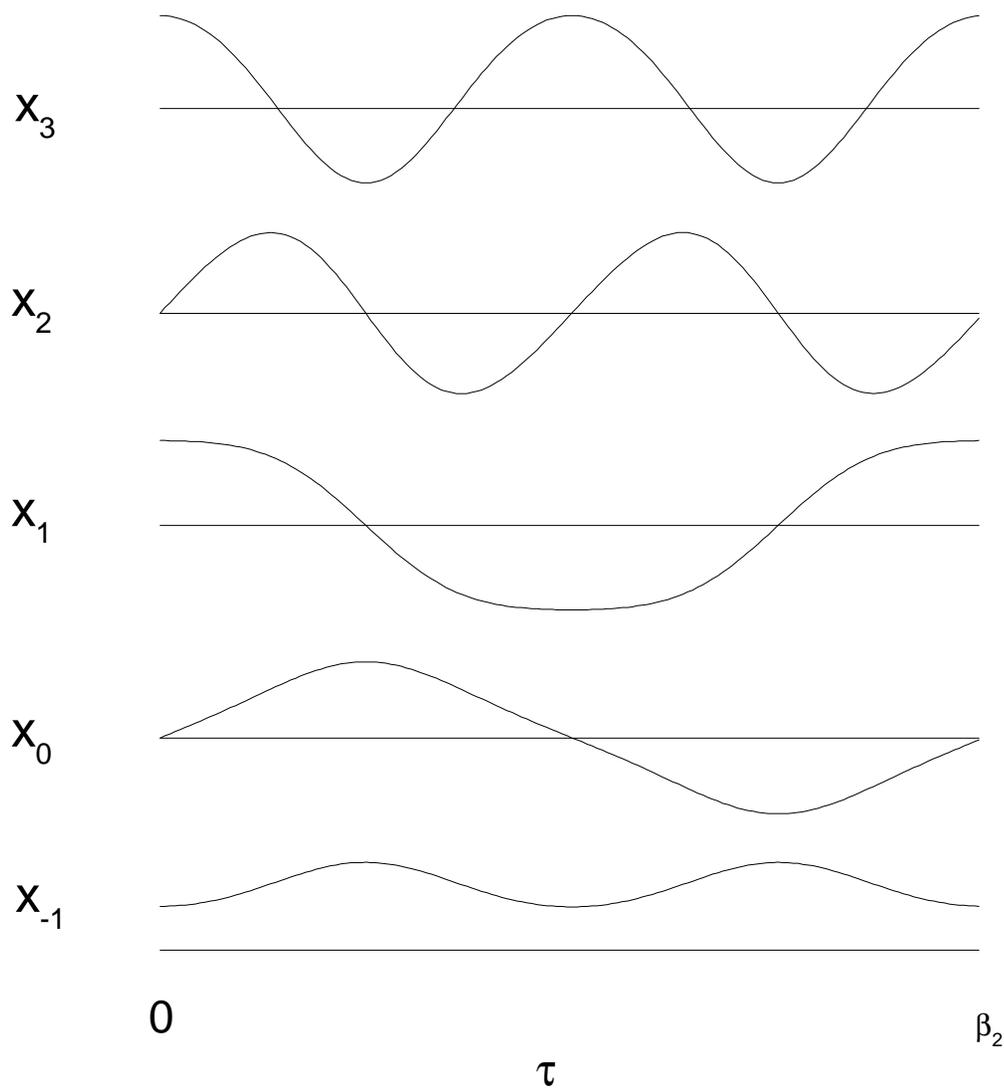
\begin{figure}
\caption{ (a) Plots of all possible solutions whose period is 
$\beta_2 > \beta_s$. $\phi_p^*$ is obtained from $\phi_p$ by changing
$\tau \rightarrow \tau - \beta_2 / 2$. 
(b) Plots of several eigenfunctions around $\phi_p$. }
\end{figure}
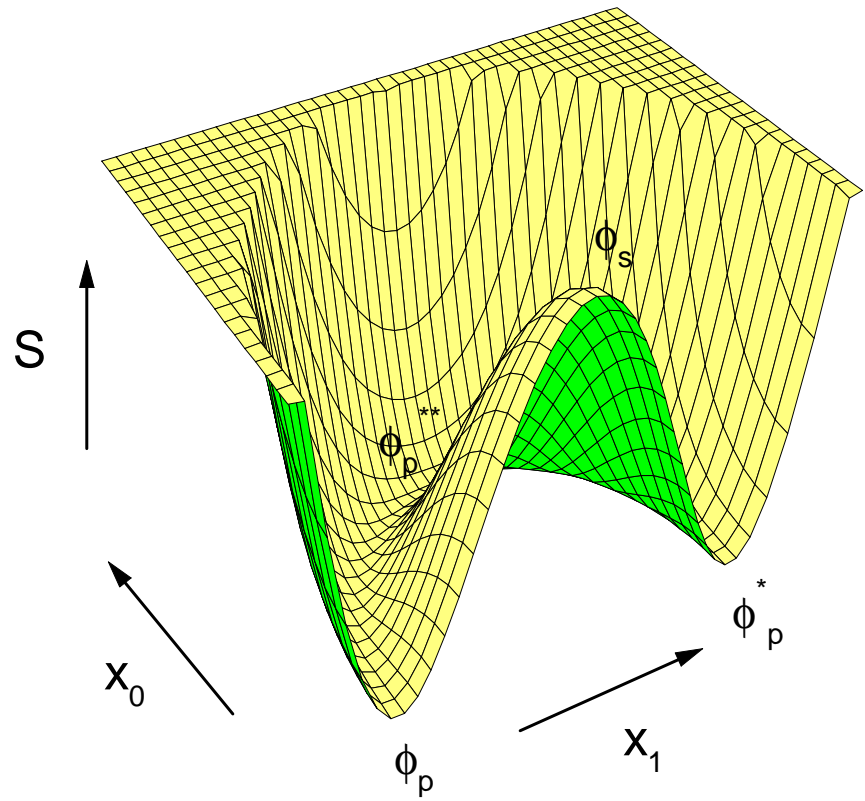
\begin{figure}
\caption{Graphical description of action manifold projected on
($x_0$,$x_1$) plane when $\beta=\beta_2 > \beta_s$.}
\end{figure}
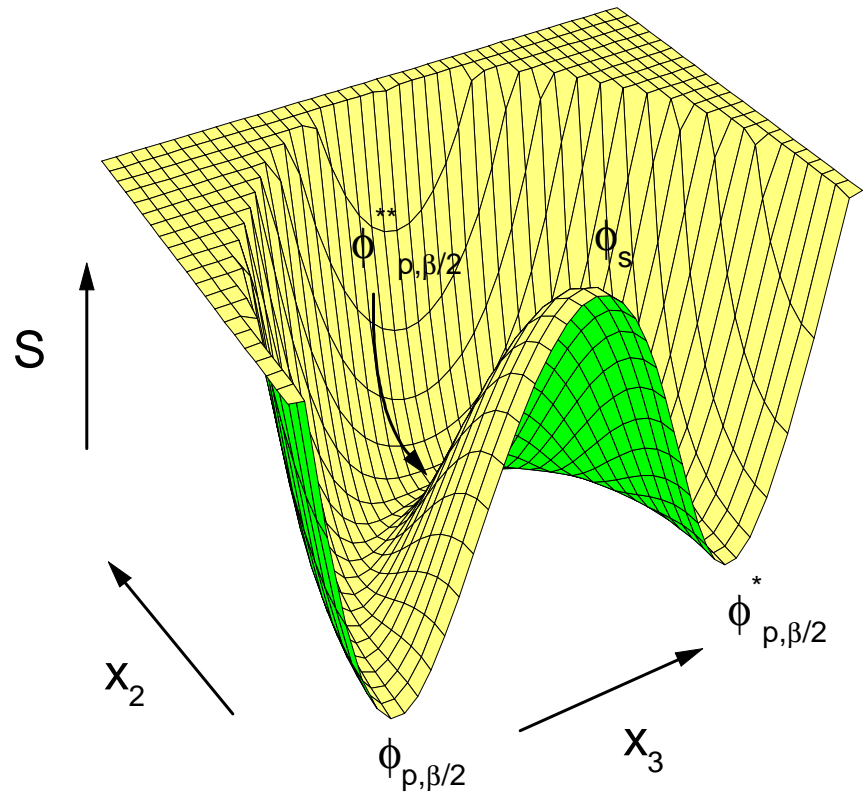
\begin{figure}
\caption{Graphical description of action manifold projected on
($x_2$,$x_3$) plane when $2\beta_s< \beta < 3\beta_s $.
}
\end{figure}
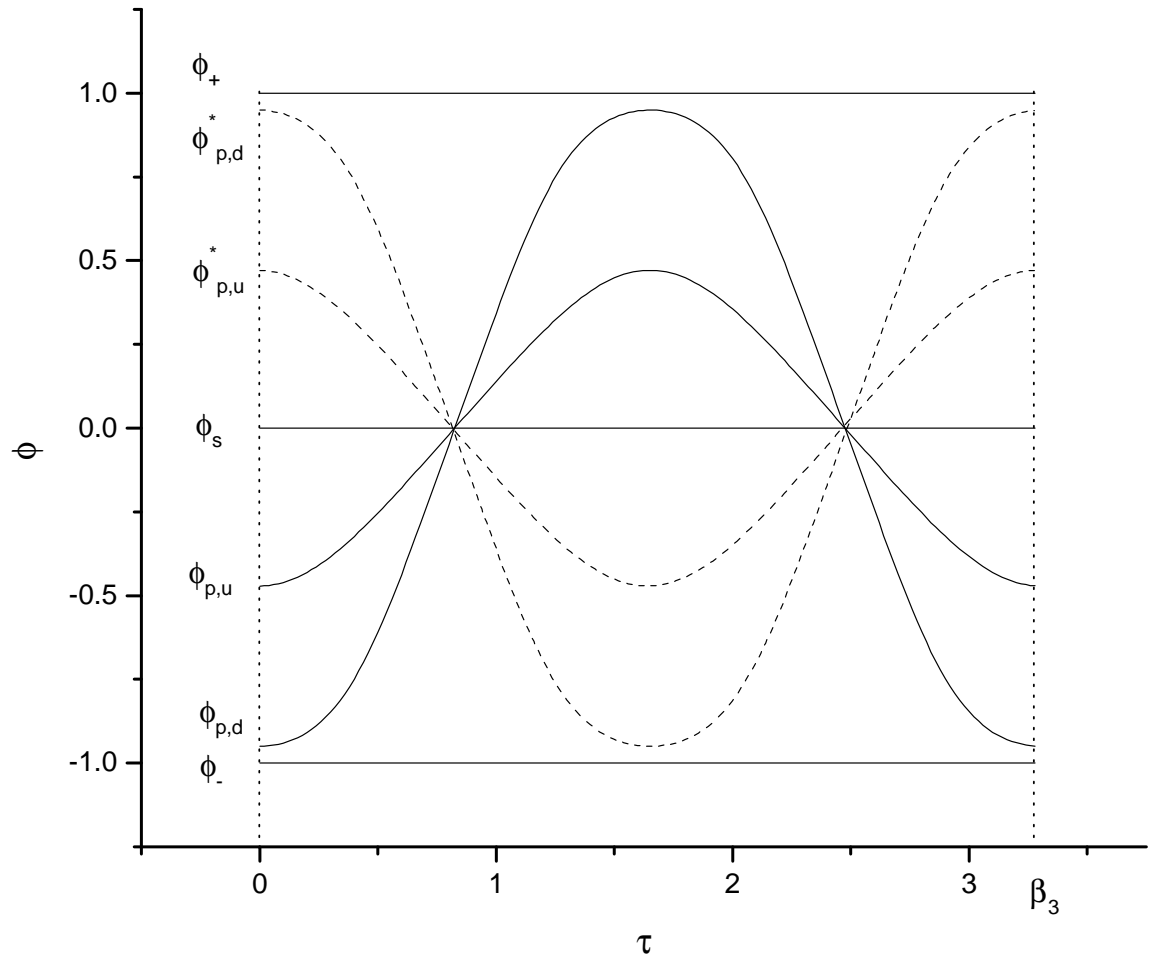
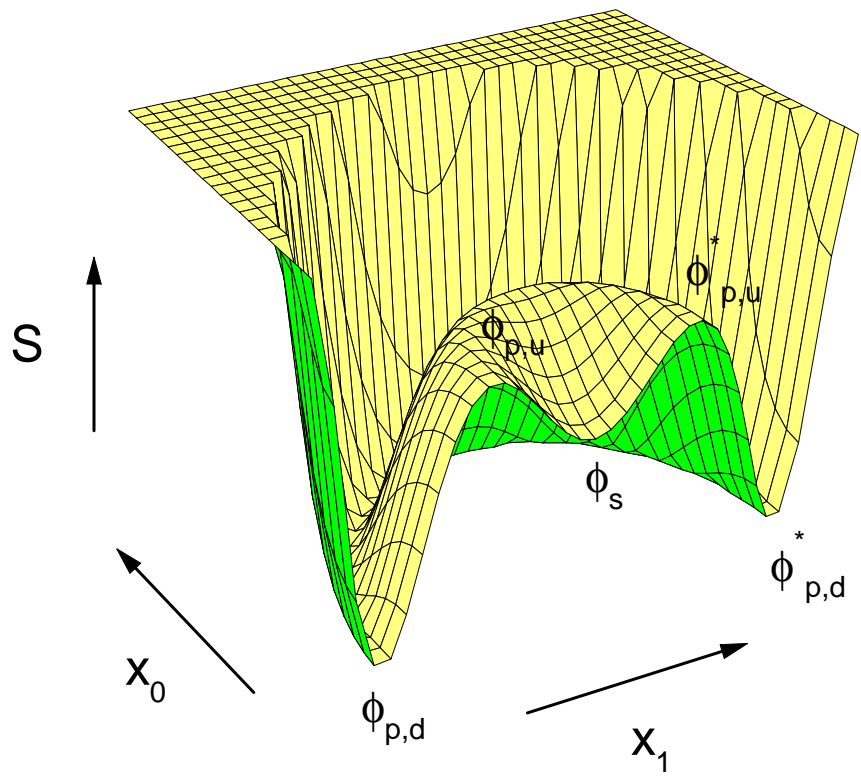
\begin{figure}
\caption{(a) Plots of all possible solutions with $\beta=\beta_3$.
(b) Graphical description of action manifold projected on
($x_0$,$x_1$) plane. 
}
\end{figure}

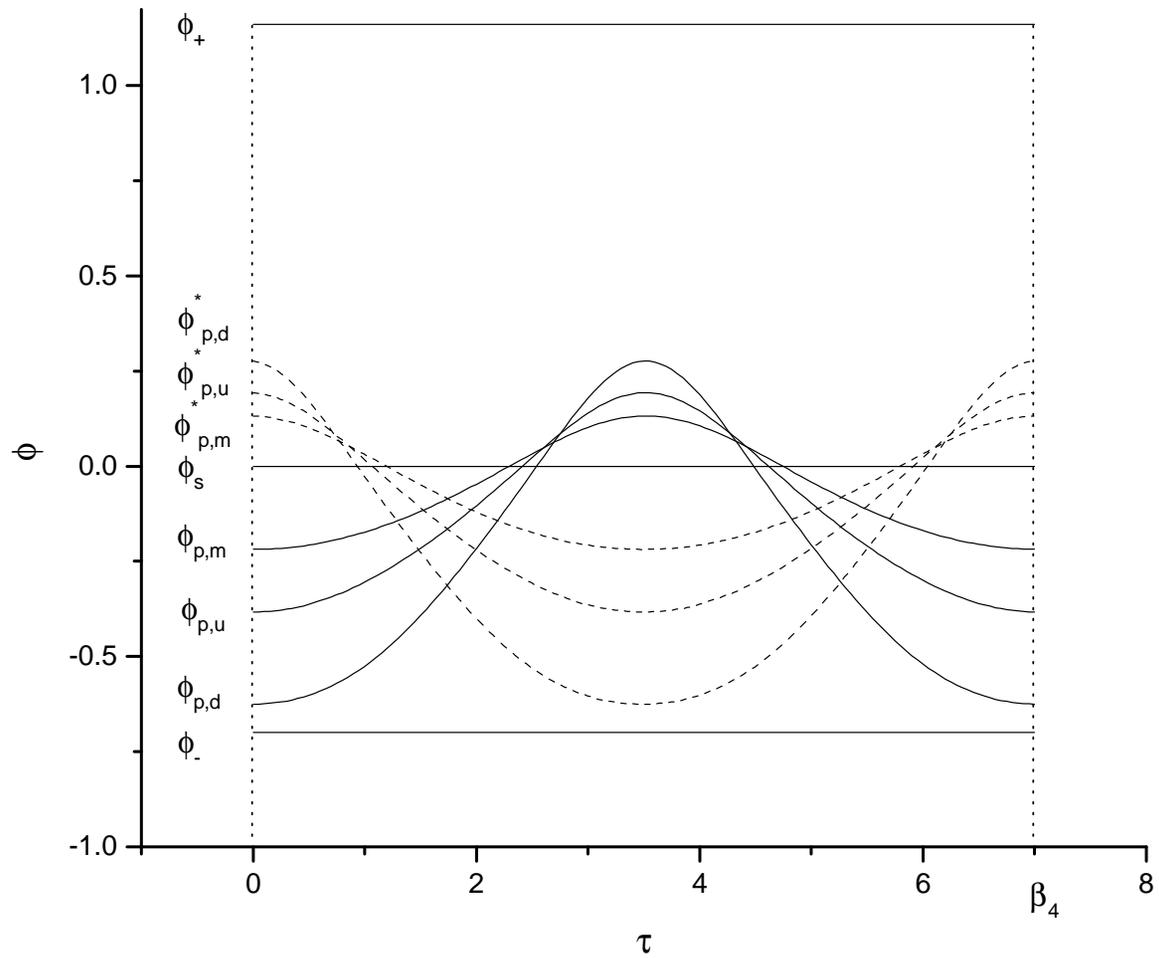
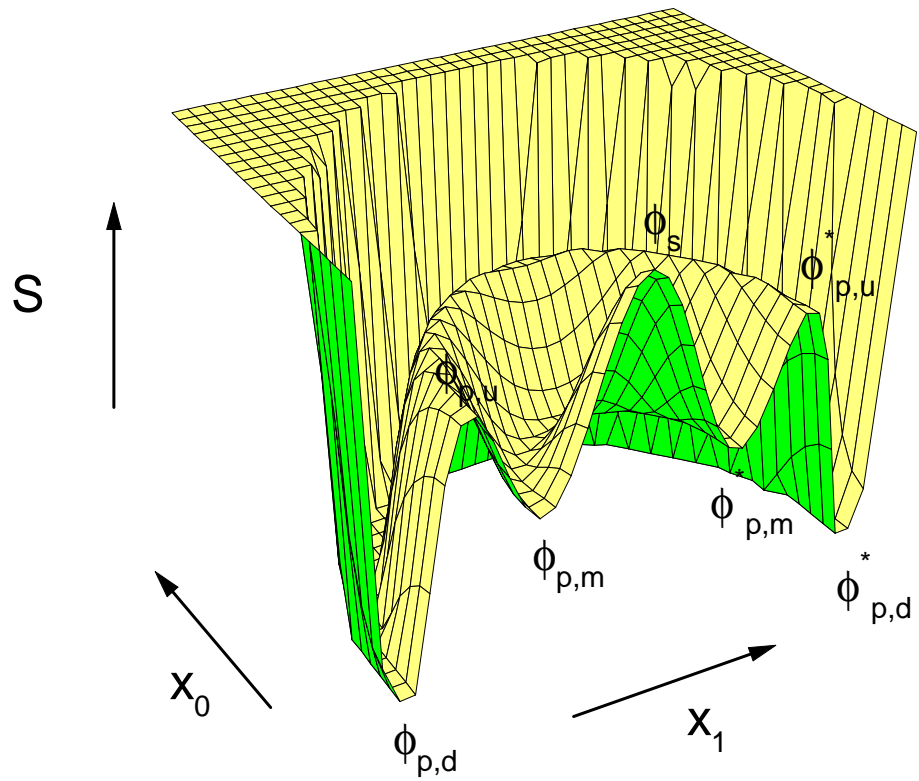
\begin{figure}
\caption{(a) Plots of all possible solutions with $\beta=\beta_4$.
(b) Graphical description of action manifold projected on
($x_0$,$x_1$) plane. 
}
\end{figure}

\end{document}